Abstract: Phobos, a moon of Mars, is below the Clarke's synchronous orbit and due to tidal interaction is losing altitude. With this altitude loss it is doomed to the fate of total destruction by direct collision with Mars. On the other hand Deimos, the second moon of Mars is in extra-synchronous orbit and almost stay put in the present orbit. The reported altitude loss of Phobos is 1.8 m per century by wikipedia and 60ft per century according to ozgate url . The reported time in which the destruction will take place is 50My and 40My respectively. The authors had proposed a planetary-satellite dynamics based on detailed study of Earth-Moon[personal communication: http://arXiv.org/abs/0805.0100 ]. Based on this planetary satellite dynamics, 2 m/century approach velocity leads to the age of Phobos to be 23 Gyrs which is physically untenable since our Solar System's age is 4.567Gyrs. Hence the present altitude loss is assumed to be 20 m per century. This leads to the age of Phobos to be 2.3Gyrs and age of Deimos to be 2.26Gyrs which is an acceptable result and from this analysis it is predicted that the travel time from the present orbital radius of 9380 km to the Martian surface at 3397 km is 10.4Myrs. Hence doomsday of Phobos is at 10.4Myrs from now.Mars Express studies have confirmed that Phobos is indeed trapped in a death spiral.

# Irregularly Shaped Satellites-Phobos & Deimos- moons of Mars, and their evolutionary history.

B. K. Sharma[1] & B.Ishwar[2]

**Abstract:** Phobos, a moon of Mars, is below the Clarke's synchronous orbit and due to tidal interaction is losing altitude. With this altitude loss it is doomed to the fate of total destruction by direct collision with Mars. On the other hand Deimos, the second moon of Mars is in extra-synchronous orbit and almost stay put in the present orbit. The reported altitude loss of Phobos is 1.8 m per century by wikipedia and 60ft per century according to ozgate url . The reported time in which the destruction will take place is 50My and 40My respectively. The authors had proposed a planetary-satellite dynamics based on detailed study of Earth-Moon[personal communication: http://arXiv.org/abs/0805.0100 ]. Based on this planetary satellite dynamics, 2 m/century approach velocity leads to the age of Phobos to be 23 Gyrs which is physically untenable since our Solar System's age is 4.567Gyrs. Hence the present altitude loss is assumed to be 20 m per century. This leads to the age of Phobos to be 2.3Gyrs and age of Deimos to be 2.26Gyrs which is an acceptable result and from this analysis it is predicted that the travel time from the present orbital radius of 9380 km to the Martian surface at 3397 km is 10.4Myrs. Hence doomsday of Phobos is at 10.4Myrs from now. Mars Express studies have confirmed that Phobos is indeed trapped in a death spiral.

In this paper we have utilized Planetary Satellite Dynamics to calculate the rate of altitude loss of Phobos which is in sub-synchronous orbit. Section 1 gives the work done on Phobos till date. Section 2 gives the

Manuscript received March 23, 2009.
  [1]Bijay Kumar Sharma is with National Institute of Technology in Electronics & Communication Department, Patna 800005,Bihar, India.(res. Phone. 091-612-2672723, Fax: 091-612-2670631; e-mail: bijay_maniari@rdiffmail.com; e-mail: bijay_maniari@hotmail.com; URL: www.geocities.com//bijay_maniari;)
[2]Bhola Ishwar is Professor Emiritus at Post Graduate Department of Mathematics, BRA Bihar University, Muzaffarpur, Bihar 842001.

planetary satellite dynamics as developed through the rigorous analysis of Earth-Moon System. Section 3 describes the different scenarios of Satellite Evolution. Section 4 gives the history of Mars-Phobos-Deimos studies, the globe-orbit parameters of Mars-Phobos-Deimos, the deduction of velocity of approach of Phobos and velocity of recession of Deimos and from this the time integral equation is set up and solved. This calculation gives the age of Phobos as well as the timetable of its doomsday. Section 5 gives the discussion and Section 6 gives the conclusions.

**Section 1. A brief history of Mars-Phobos-Deimos. (Sangdeev & Zakharov 1989)**

Phobos and Deimos are the two moons of Mars. They were discovered by Asaph Hall in 1877. The history of the studies of Mars and its moons are given in Table 1.

Table 1. History of the studies of Mars and its moons.

| Year | Person or Spacecraft | Work done. |
|---|---|---|
| 1659 | Christian Hugens | Drew the first sketch of the dark and bright side of Mars. |
| 1780 | William Herschel | Noted thin Martian Atmosphere. |
| 1877 | Giovanni Schiapaprelli | Drew first detailed map of Martian surface. |
| 1900 | Percival Lowell | Used Lowell Telescope to make drawing of the canals on Martian Surface. |
| 1965 | Mariner 4 | Beamed back 20 photos from first flyby of Mars. |
| 1971 | Mariner 9 | Sent back 7300 images from first ever orbital mission. An interlocking grid covered Video Frame 4209-75 was one of the images. |
| 1976 | Viking 1 & 2 | First probes to land on Martian Surface and photograph the terrain. |
| July 7,1988 | Phobos 1 | It failed enroute. On September 2, 1988, it lost its lock on Sun due to software glitch and hence it lost its power source. |
| July 12, 1988 | Phobos 2 | It became Mars Orbiter on January 29,1989, and sent 38 images with a resolution of 40m. It has gathered data on Sun, Interplanetary Medium, Mars & Phobos. A base station and a Mars rover |

|             |                                   | was to be released but the all contact was lost on March 29,1989. One of the images are similar to Frame 4209-75 sent by Mariner 9. |
|-------------|-----------------------------------|------------------------------------------------------------------------------------------------------------------------------------|
| 1998        | Mars Global Surveyor              | It is mapping the whole surface of Mars                                                                                            |
| 2002        | Mars Odyssey                      | It took night time I.R. pictures of Martian Crater called Hyataspis Chaos.                                                         |
| 2003        | European Space Agency Mars Express | (1) It has revealed the volcanic past of Mars; (2) Icy Promethei Planum , the icy south pole of Mars , has been photographed; (3) In 2008 Atmosphere stripping on Mars and Venus are being simultaneously studied by Mars Express and Venus Express. |
| 27 May,2008 | Phoenix                           | Soft Landed on North Pole of Mars in search of extraterrestrial life                                                               |

Grey Phobos and Deimos are quite unlike ruddy, pink-skied planet Mars. The two natural satellites are pitted and like drought-state potato. Their surfaces are seared by meteorites and raked by solar wind. They have much lighter density and are probably formed of carbonaceous chronditic material in outer part of the asteroid belt (Burns 1978). The central force of these lilliputian natural satellites are weak hence the constituent materials have not undergone compaction. These natural satellites have escaped the deeper trauma of heating and inner shifting that have occurred in the formation of Planets.

There are numerous asteroid families in the outer realm of Asteroid Belt which are in inclined orbits with angle of inclination 20º or more to the Ecliptic. These asteroids are of carbonaceous chondritic composition and likely to be perturbed by Jupiter into Mars-crossing orbits. These asteroids are potential candidates for capture by Mars (Lambeck 1979).

It is widely believed that the two irregular shaped satellites( Chang & Lorre, 2000; Thomas 1979; Thomas,1989; Duxbury 1974; Stook & Keller 1990) of Mars are captured asteroids. Under the influence of Jupiter, some asteroids in remote past might have been catapulted into the inner part of the Solar System. It could also be that 2:1 Saturn-Jupiter Mean Motion Resonance crossing at 300My after the birth of Jupiter might have catapulted Neptune into the outer part of the Solar System stirring up the Oort Cloud (Theoretical Formulation of the origin of cataclysmic late heavy bombardment era based on the new perspective of birth & evolution of solar systems. http://arXiv.org/abs/0807.5903 ) . This might have started a rain of comets into the inner part of the Solar System. This rain of Comets in turn might have disturbed the asteroid belt. The disturbance in Asteroid belt might have started the Late Heavy Bombardment (LHB) Era between 500My to 700 My after the birth of Jupiter. As a result of this LHB Era a large number of asteroids might have been catapulted in the inner part of the Solar System. In course of this journey the two asteroids may have been captured by Mars about 2.5 Gya. Infact there are recent evidences by the study of the dark side of our Moon that events associated with LHB persisted up to 2.5Gya (Neumann & Mazarico 2009). GASPRA (Asteroid 951 discovered in 1916 by Russian Astronomer Grigorii Nikalaevich Noujmin,1886-1946) is very similar to these Martian satellites in its general features but still resides in the Asteroid Belt. It is a S-class asteroid and has craters and linear grooves similar to those found on Phobos(A Dictionary of Astronomy 1997). This asteroid was the first to be studied in detail by Galileo Probe in 1991 when in planet flyby mission within 1600km it took its photographs. Gaspra is irregularly shaped object with dimensions of 18.2×10.4×8.8km and is probably a fragment of much larger body. Its spin period is 7.04hrs and orbital period is 3.28 years. Semi-major axis a= 2.21AU, perihelion=1.83AU , aphelion = 2.59AU and inclination i= 4º1'.

**Section 2. Earth-Moon System Revisited.**

From George Howard Darwin's time it is recognized that planets raise body tides in their respective natural satellites and natural satellites raise body tides in their respective planets. It is also recognized that planets and satellites are anelastic bodies. Hence tidal deformation leads to dissipation of energy called tidal dissipation. By assuming different Love Numbers (Q parameter) different rate of tidal dissipation can be incorporated in the tidal interaction. Tidal interaction inevitably leads to tidal drag on the primary object if the satellite is above synchronous orbit or tidal acceleration of the primary object if the satellite is in sub-synchronous orbit and zero tidal interaction if the two bodies are tidally interlocked. If there is a tidal interaction then satellite orbits will be an evolving orbit like an expanding spiral or contracting spiral (Burns 1978; Lambeck 1979; Szeto 1983;Sharma,Ishwar & Rangesh 2009). Satellites like our Moon and Deimos which are in extra-synchronous orbit are in expanding spiral orbit whereas Phobos which is in sub-synchronous orbit is in contracting spiral orbit whereas as Charon (moon of Pluto) which is exactly in mutually interlocked tidally synchronous orbit is in stationary circular orbit. Assuming Q parameter, Darwin-Kaula formulation results in secular evolution equations. These secular evolution equations can be integrated into the past to see how it has evolved since its inception. This integration can be extrapolated into future to see where and when eventually it terminates. It is estimated that Phobos from the present orbit of 9830km from the center of Mars will spirally collapse to an orbital radius of 3397km (the martian surface) in about 100My (Burns 1978). This paper estimates 10.4My for the same orbital collapse. Other researchers have made a wide range of assumptions regarding models of dissipation by anelastic tidal deformation within Mars and satellites to test the Capture Hypothesis(Szeto 1983). Szeto has proposed that Capture would have led to collision but no collision seems to have occurred in last 1.5Gy. Also collision could not have resulted into near circular orbit of Deimos though it could have led to gravitationally runaway orbit of Phobos. Hence by general consensus of the older researchers ,the capture origin is discarded(Goldreich 1963 ; Singer, 1970; Lambeck1979; Szeto 1983). But our analysis points out that Capture Model is more suitable for explaining the presence of irregularly shaped Phobos and Deimos in orbit around Mars.

It is well established that our Moon is receding from the Earth at the rate of 3.7cm/yr ( Dickey 1994) . It is also well established that it is spiraling out until it will get into a geosynchronous orbit or Clarke's orbit (Kaula & Harris 1975). In this futuristic orbit it will be orbiting in 47 days. But what has not been known commonly that Roche's Limit (Ida et al 1997) of 18,000 Km lies just beyond the inner Geosynchronous Orbit. At the inner Geosynchronous Orbit, length of month = length of day= 5 hours where both Earth and Moon are tidally interlocked [personal communication http://arXiv.org/abs/0805.0010 ]. At that point of accretion from the circumterrestrial impact generated debris, the fully formed Moon experienced a Gravitational Sling Shot Effect (Cook 2005, Dukla et al 2004, Epstein 2005, Jones 2005) which launched it on an outward expanding spiral path. Gravitational Sling Shot is termed as Planet Fly-by-Gravity-Assist maneuver which is routinely used to boost mission spacecrafts to explore the farthest reach of our Solar System. Gravitational sling Shot creates an impulsive torque which gave the orbiting Moon its extra rotational energy with which it continues to spiral out and climb up the potential well created by a much heavier Earth.

It has been proposed by us that since all natural satellites are either in spirally collapsing orbits as Phobos is or in spirally expanding orbit as our Moon and Deimos are, therefore all natural satellites have been born at the inner Clarke's Orbit , experienced Gravitational Sling Shot effect and either they have been launched on a collapsing gravitational runaway sub-synchronous orbit or on expanding extra-synchronous orbit. This has been observed by Lambeck(1979). He backward extrapolated the evolutionary orbits of Phobos and Deimos and found that both lead to the same spatial region of origin which we assert is inner Clarke's Orbit. Our orbital evolution is constrained only by the age of the satellites. We assume that if the age is known, we know the transit time from '$a_{G1}$' (inner Clarke's Orbit) to 'a' ( the present orbital radius). This information is enough to determine the dynamical evolutionary equation of the secondary body. In case of the research done till now, a wide range of assumptions concerning models of dissipation by anelastic tidal deformation have to be considered before arriving at a realistic evolutionary orbit (Goldreich1963; Singer 1970; Kurt Lambeck1979; Szeto 1983) .

**Section 3. Different scenarios of evolution of Planetary Satellites.**

It is Gravitational Sling Shot impulsive torque which enabled Charon to spiral out from the inner Clarke's Orbit to outer Clarke's Orbit where it is in stable equilibrium tidally interlocked with Pluto (Sharma & Ishwar 2004). If our Moon had fallen short of the inner Geosynchronous Orbit it would have been launched on a gravitational runaway collapsing spiral path to its certain doom. The Phobos is launched on precisely such an inward gravitational runaway collapsing spiral path because it orbits below the inner Clarke's Orbit. It is losing its altitude at the rate of 60ft per century which comes out to be 18.29m/century [www.ozgate.com/infobytes/mars_views.htm]. Wikipedia gives 1.8m/century. This paper arrives at 20 m/century of altitude loss by applying planetary satellite dynamics as developed by the authors [personal communication http://arXiv.org/abs/0805.1454 ] and by assuming an age of 2.3Gy for Phobos.

**Section 4. Mars-Phobos-Deimos System Analysis.**

**Section 4.1 Age of Phobos.**

Phobos is the least reflective body in our Solar System largely constituted of carbonaceous chondrite material called Type-C asteroids (lying in outer part of the Asteroid Belt) and captured early in Solar history. Mars Express has revealed that it is relatively red in colour resembling D-Type Asteroids (lying at the outer edge of the main Asteroid Belt). Phobos is thought to be made of ultra primitive material containing carbon as well as ice but it has experienced even less geo-chemical processing than many carbonaceous chondrites. Hence Phobos date of capture is kept at more than 2.5 Gy. We will assume the date of capture at 2.3Gy.

**Section 4.2 The Kinematics of Mars-Phobos & Mars-Deimos.**

The inner Clarke's synchronous orbit of Phobos is 20,400 km. Phobos at an orbital radius of 9380 km (about 6000km above the Martian surface) and with an orbital period of 7 hrs 39mins and Mars spinning at 1.026days period is causing Phobos to be gradually being drawn inward. Altitudinal loss rate is 1.8 m/century as given by Wikipedia and 60 ft per century as given by www.ozgate.com/infobytes/mars_views.htm .It is estimated that within 50 Myrs to 40 Myrs Phobos will crash into Mars (Duxbury 2007, Wikipedia, Ozgate). Our analysis or the New Perspective gives the crash time as 10.4My from now. The altitude loss is at the rate of 20 cm per year or 20 m per century assuming an age of Phobos as 2.3Gyrs

It has been shown [personal communication: http://arXiv.org/abs/0805.0100 ] that:

$\omega/\Omega = P_1/P_2 = LOM/LOD = E \cdot a^{3/2} - F \cdot a^2$     (1)

where $E = J_T/(BC)$;

$F = (m/(1+m/M))(1/C)$;

C = moment of inertia of Mars around its rotation axis.

$B = \sqrt{(GM(1+m/M))} = \sqrt{(G(m+M))}$;

$J_T = C\omega + (m/(1+m/M))B\sqrt{a}$ ;

$\omega = (2\pi/P_2)$ = planet's angular spin velocity ;

$\Omega = (2\pi/P_1)$ = satellite's angular orbital velocity;

**Table 2. Globe and Orbit Parameters of the Mars-Phobos-Deimos system (Chaisson et al 1998, Hannu et al 2003, Moore 2002)**

|  | M(kg) Mass of Mars | $R_M$ (m) radius of Mars | C (kg-m²) | a (m) | $P_1$ (solar d) | $P_2$ (solar d) | m(kg) Mass of Phobos | $\rho_M$ (gm/cc) density of Mars | $\rho_m$ (gm/c.c.) density of Phobos |
|---|---|---|---|---|---|---|---|---|---|
| Mars-Phobos | $6.4191 \times 10^{23}$ | $3.397 \times 10^6$ | $2.9634 \times 10^{36}$ | $9.38 \times 10^6$ | 0.319 | 1.026 | $1.1 \times 10^{19}$ | 3.93 | 2.0 |
| Mars-Deimos | same | $1.15 \times 10^6$ | same | $2.346 \times 10^7$ | 1.262 | same | $1.8 \times 10^{18}$ |  | 2.7 |

C= Moment of Inertia around the spin axis of the Planet Mars $=(2/5)MR_M^2$;

a = semi-major axis of the moon;

$P_1$= satellite's orbital period;

$P_2$= planet's spin period;

$a_R$= Roche's radius = $2.456(\rho_M/\rho_m)^{1/3}R_M$ =10,426km

Roche's Zone= 0.8 to 1.35 of $a_R$ = 8,340km to 14,075km

**Table.3. Parameters E, F, $a_{G1}$, $a_{G2}$, $a_R$, lom/lod$|_{cal}$ & lom/lod$|_{obs}$ of Mars-Phobos &Mars-Deimos.**

|     | B | $J_{spin}$ Mars | $J_{orb}$ | $J_T$ | E | F | $a_{G1}$ (m) | $a_{G2}$ (m) | $a_R$ (m) | $(\omega/\Omega)_{obs}$ | $(\omega\Omega)_{cal}$ |
|-----|---|-----------------|-----------|-------|---|---|--------------|--------------|-----------|-------------------------|------------------------|
| M-P | 6.54^ ×10⁶ | 2.101 ×10³² | 2.19 ×10²⁶ | 2.1005 ×10³² | 1.083 ×10⁻¹¹ | 3.683 ×10⁻²¹ | 2.04 ×10⁷ | 8.65 ×10¹⁸ | 10.4 ×10⁶ | 0.3109 | 0.311 |
| M-D | 6.54^ ×10⁶ | 2.101 ×10³² | 7.53 ×10²⁵ | 2.1005 ×10³² | 1.083 ×10⁻¹¹ | 8.016 ×10⁻²² | 2.04 ×10⁷ | 1.83 ×10²⁰ | 9.41 ×10⁶ | 1.2305 | 1.2306 |

Roots of LOM/LOD=1 give the two geosynchronous orbits or the two Clarke's Orbits $a_{G1}$ and $a_{G2}$.

Using Mathematica the two roots or the two Clark's orbits of Phobos are:

$a_{G1}$ = Inner Clarke's Orbit= $2.04 \times 10^7$ m = 20,400Km;

$a_{G2}$ = Outer Clarke's Orbit= $8.65 \times 10^{18}$ m = $8.65 \times 10^{15}$ Km;

Two Clarke's orbits of Deimos are:

$a_{G1}$ = Inner Clarke's Orbit= $2.04 \times 10^7$ m = 20,400Km;

$a_{G2}$ = Outer Clarke's Orbit= $8.65 \times 10^{18}$ m = $1.83 \times 10^{17}$ Km;

As we see that inner synchronous orbits are the same for both the satellites namely 20,400km. In our perspective we assume that the inception of the satellite takes place at inner synchronous or at inner Clarke's Orbit. This implies that both satellites proceeded on their spiral orbit from $a_{G1}$ : Phobos on a contracting spiral orbit or gravitationally runaway collapsing orbit/death spiral orbit and Deimos on an outward expanding spiral orbit except that the expanding spiral orbit of Deimos is not evident because the time constant of evolution is inordinately large as we will see later in this section and as has been verified by Lambeck(1979) . In fact backward extrapolated orbits of both satellites lead to a similar region of origin within the age of Solar System by other researchers also (Lambeck 1979).

For all practical purposes $a_{G2}$ is infinity for both Phobos and Deimos . If Phobos had tumbled beyond $a_{G1}$ it would never evolve out of $a_{G1}$ . This is because there is a term Time Constant of Evolution which is inversely proportional to the mass ratio of satellite to planet. If the mass ratio is infinitesimally small then Time Constant is infinite and satellite remains stay put in the first Clarke's orbit. But if it is significant then satellite very rapidly evolves from the inner to outer Clarke's orbits as is the case with Charon, a satellite of Pluto, as is the case with our Moon (Sharma,Ishwar,Rangesh 2009) and as already seen in previous papers (Sharma & Ishwar 2004, Sharma & Ishwar 2004a, Sharma & Ishwar 2004b) . Deimos point of inception is 20,400km and present orbital radius is at 23,460 km that amounts to 1.5% orbital expansion in 2.3Gy. The Time Constant of evolution for Phobos and Deimos are $1.23 \times 10^{12}$Gy and $2.07424 \times 10^{13}$Gy which is infinity for practical purposes . The man made satellites around Earth's geo-synchronous orbit is a point mass in comparison to Earth hence mass ratio is infinitesimal therefore it is almost in a non-evolutionary orbit. For man-made satellites there is only one geo-synchronous orbit at 36,000 km above the equator. The other is at infinity. Man-made Geo-synchronous Satellites remain stay-put at 36,000 km orbit. They do not evolve!

**Section 4.3 Derivation of Velocity of Approach and setting up the time Integral Evolution Equation.**

All the relevant globe-orbit parameters are given in Table 2. As seen from the Table 3, lom/lod by calculation and observation are the same. So we can say that lom/lod equation is correctly derived.

**Table 4. Tabulation of sling-shot point x1, Gravitational Resonance point(x2) and Structure Factor parameters M and K.**

|        | $X_1$(m)         | Lom/lod$|_{x1}$ | $X_2$(m)             | M   | K(N-m$^{M+1}$)       |
|--------|------------------|----------------|----------------------|-----|----------------------|
| Phobos | $2.37\times10^7$ | 1.25           | $3.24291\times10^7$  | 3.5 | $2.72108\times10^{38}$ |
| Deimos | $2.37\times10^7$ | 1.25           | $3.24291\times10^7$  | 3.5 | $5.65\times10^{37}$  |

**Table 5. Tabulation of the kinetics, time constant of evolution($\tau = (a_{G2}-a_{G1})/V_{max}$), transit time from $a_{G1}$ to $a_{present}$=Age of the natural satellite and the evolution factor $€=(a-a_{G1})/(a_{G2}-a_{G1})$ of the natural satellites.**

|        | Radial Acceleration Observed* | $V_{max}$ m/yr | $V_{present}$ m/yr | €                    | $\tau$ (Gyr)        | Age (Gyr) |
|--------|-------------------------------|----------------|--------------------|----------------------|---------------------|-----------|
| Phobos | $843.32\times10^{-6}$         | 0.007          | -0.199             | $-1.28\times10^{-12}$ | $1.235\times10^{12}$ | 2.32      |
| Deimos | $106.8\times10^{-6}$          | 0.009          | 0.0054             | $-1.6605\times10^{-14}$ | $2.074\times10^{13}$ | 2.26      |

*Radial Acceleration=$[\Omega^2 a_{present}/(1+m1/m0) - GM/a_{present}^2]$

Detailed derivation of LOM/LOD by calculation and the derivation of radial velocity of Recession/Approach is given in our personal communication http://arXiv.org/abs/0805.1454 :

We obtain the present rate of Approach for Phobos or present rate of Altitudinal Loss:

-0.1992672m/yr = -19.92672 m/century for Phobos.

This comes out to be 20 cm/year or 20 m/century assuming an age of Phobos to be 2.3Gy. The age of Phobos at 2.3Gy implies that Phobos took the transit time of 2.3Gy from the point of capture to the present position.

For determining the transit time from the point of capture to the present position, we will have to solve the time integral Equation [ personal communication ibid].

With 20m/century rate of approach , the Age of Phobos comes to be 2.3Gyrs. If present rate of approach is adopted to be 2m/century the age of Phobos comes to be 23Gyrs which is physically untenable since our Solar System is 4.567Gyrs old (Toubol et al 2007, Stevenson 2008). Hence 20m/century rate of altitudinal loss is the valid magnitude in our planetary satellite model.

### Section 4.4 Determination of the Doomsday.

To determine the time of DOOMSDAY, time integral equation [personal communication ibid] will have to be solved within the limits of $9.38 \times 10^6$m to $3.4 \times 10^6$m (globe radius of Mars). The result is 10.4Myrs.That is 10.4Myrs from now Phobos will be destroyed through head on collision with Mars..

But even before head on collision takes place , it is asserted that as soon as Phobos enters 7000km Roche's zone(Ozgate URL) above the center of Mars the primary tides will smash it and convert it into annular ring of dust which will eventually spiral into Mars.

But our analysis says that Roche's Zone lies within 8000km to 14,000km. Hence the question of Phobos being pulverized by primary tides does not arise. This is because Phobos is a captured asteroid with high tensile strength though it lacks compaction hence primary tides cannot pulverize it.

### Section 5. Discussion (www.esa.int/esaMI/Mars_Express/SEM21TVJD1E_O.html ).

Since Wikipedia and Ozgate are giving conflicting data on rate of altitudinal loss. The researchers who worked out dynamic evolutionary equations have not given any result on altitudinal loss of Phobos. Burns(1978) has given the estimate of the doomsday as 100My which seems to be in complete conflict with our result of 10.4My. Our result seems to be realistic as Phobos is in a gravitational runaway collapsing orbit. The actual measurement only can establish if the above analysis is correct.

Mars Express was inserted in Mars Orbit on 25$^{th}$ December 2003. Its first mission was completed on 31$^{st}$ October 2007. The mission got a second extension up to May 2009. Mars Express has clearly detected Phobos to be in death spiral slowly orbiting towards Mars surface. It has also been detected that Phobos is experiencing increased orbital speed due to inward secular acceleration. The orbit will be studied in more detail over the lifetime of Mars Express.

**Section 6. Conclusion.**

This analysis remains inconclusive as far as the validity of the Gravitational Sling Shot approach is concerned because we do not have an authoritative record of Phobos Altitutdinal rate of loss . Mars Express space mission will lead to a better understanding of death spiral in which Phobos is presently trapped. There could be special missions with space craft equipped with Radar and carrying out Mars Laser Ranging Experiments to get an authoritative record of rate of altitudinal loss. This experiment could be carried out from Mars itself by sounding Phobos.

**Acknowledgment:** The Author acknowledges AICTE financial support in form of 8017/RDII/MOD/DEG(222)/98-99 under MODROBS Scheme.

**Author Information:** Reprints and permissions information is available at npg.nature.com/reprintsandpermissions. The authors declare no competing financial interests. Correspondence and requests for materials should be addressed to B.K.Sharma(: bijay_maniari@rediffmail.com)

**Bijay K. Sharma** (Fellow of IETE-1985, Member IEEE-2007), Date of Birth: 01/07/1946, Place of Birth: Bela, District Sitamarhi, Bihar, India.B.Tech(Hons) 62-67,Electronics & Communication Department from Indian Institute of Technology,Kharagpur, India. M.S., 69-70, Electrical Engineering with specialization in Microelectronics from Stanford University, U.S.A.; Ph.D., 70-72, Electrical Engineering with specialization in Bio-Electronics from University of Maryland, U.S.A. Currently the author is working towards his D.Sc. degree in the field of Applied Mathematics, namely Celestial Mechanics, at BRA Bihar University, Muzaffarpur, Bihar, India.
       He worked as Junior Research Fellow at Central Electronics Engineering Research Institute, Pilani, Rajasthan, in the period 67-69. In this period he developed Tolansky Interferometer for measuring thin film thickness of the order of 200Å and Nichrome Resistance of 200 ohm/square sheet resistivity with and temperature coefficient of resistance 200p.p.m per degree. In 69-72, he completed his higher studies in U.S.A.. After returning to India he worked as Pool Officer in Electrical Engineering Department of Bihar College of Engineering, Patna, Bihar, India, during the period 72-73. This was

sponsored by Council of Scientific and Industrial Research. From 73-80, he was engaged in grass root activism in the villages of Bihar. From 80-84, he served as a lecturer in E & ECE Department of I.I.T., Kharagpur. 84-85, he served as Assistant Professor in Electronics & Electrical Engineering Depaertment at Birla Institute of Technology & Science. 85-86, he served as Assistant Professor and Head of the Department in Electronics Engineering Department of Institute of Engineering & Technology, Lucknow, U.P. From 86 to 97, he was engaged in rural construction activities at his village home in Maniari, Muzaffarpur, Bihar, India. From Dec 97 to date he has been actively pursuing R&D in the field of Microelectronics in Electronics & Communication Engineering Department in National Institute of Technology, Patna. Presently he is in the post of Assistant Professor and Head of the Department. He has been actively engaged with the study of Planetary Satellites in his work for D.Sc. under Prof. Bhola Ishwar.

    Dr. Sharma in course of his R & D work has written a series of papers on Universal Hybrid $-\pi$ model which will go a long way in developing and establishing an accurate compact model for analog circuit and system simulation using BJT. In course of his D.Sc. Dr. Sharma has proposed a new theory of Solar System's birth and evolution. With the discovery of exo-planets and their inventory increasing every day both in number as well as in diversity, Celestial Physicists are at complete loss in terms of a coherent theory which can consistently explain the diversity of exoplanets. Dr. Sharma's theory fills the gap. Also other people's work is corroborating the work of Dr. Sharma.